\documentclass[aps,twocolumn,prl,showpacs,superscriptaddress,groupedaddress]{revtex4}

\usepackage{dcolumn}% Align table columns on decimal point
\usepackage{bm}% bold math
\usepackage{color}

\def\ltape{\hbox{\ $<$\hskip -8pt\raise -4pt\hbox{$\sim$}\ }}
\def\gtape{\hbox{\ $>$\hskip -8pt\raise -4pt\hbox{$\sim$}\ }}

\pdfoutput=1

   \ifx\pdfoutput\undefined
   \usepackage[dvips]{graphicx}
   \else
   \usepackage[pdftex]{graphicx}
   \fi
   \usepackage{epstopdf}
   \DeclareGraphicsExtensions{.pdf,.gif,.jpg,.eps,.png}
  
   \usepackage{amsmath}
  \usepackage{amssymb}
\begin{document}

%\preprint{{\it Physical Review Letters}} 

\title{A model of global magnetic reconnection rate in relativistic collisionless plasmas }
%\title{Do Dispersive Waves cause Fast Collisionless Magnetic Reconnection?}

\author{Yi-Hsin~Liu}
\affiliation{NASA-Goddard Space Flight Center, Greenbelt, MD 20771}
\author{Michael~Hesse}
\affiliation{NASA-Goddard Space Flight Center, Greenbelt, MD 20771}
\author{Fan~Guo}
\affiliation{Los Alamos National Laboratory, Los Alamos, NM 87545}
\author{William~Daughton}
\affiliation{Los Alamos National Laboratory, Los Alamos, NM 87545}
\author{Hui~Li}
\affiliation{Los Alamos National Laboratory, Los Alamos, NM 87545}

\date{\today}

\begin{abstract}
%A model of the global reconnection rate in relativistic collisionless magnetic reconnection is developed. Based on the force balance upstream and downstream of the diffusion region, we explain why the global rate remains $\lesssim 0.2$ even when the local rate goes up to $\sim O(1)$ and the inflow speed approaches the speed of light in strongly magnetized plasmas. This study suggests that the magnitude of the reconnection electric field is bounded by $\sim20\%$ of the reconnecting component of magnetic field in the extremely relativistic limit. The model is general and can be applied to magnetic reconnection under widely different circumstances. 

A model of global magnetic reconnection rate in relativistic collisionless plasmas is developed and validated by the fully kinetic simulation. Through considering the force balance at the upstream and downstream of the diffusion region, we show that the global rate is bounded by a value $\sim 0.3$ even when the local rate goes up to $\sim O(1)$ and the local inflow speed approaches the speed of light in strongly magnetized plasmas. The derived model is general and can be applied to magnetic reconnection under widely different circumstances.

\end{abstract}


\maketitle

%{\color{red} \it Introduction--}
{\it Introduction--}
Magnetic fields often serve as the major energy reservoirs in high energy astrophysical systems, such as pulsar wind nebulae \cite{coroniti90a,arons12a,lyubarsky01a,devore15a}, gamma-ray bursters \cite{thompson94a,zhangB11a,mckinney12a} and jets from active galactic nuclei \cite{beckwith08a,giannios10a,jaroschek04a}, where relativistic cosmic rays and gamma rays of energies up to TeV are generated explosively \cite{abdo11a,bottcher13a}. Among the proposed physics processes (e.g.,\cite{sironi15a,uhm14a,zweibel09a}) that could unleash the magnetic energy, magnetic reconnection is considered to be a promising mechanism. 
%Magnetic reconnection not only reorganizes the topology of magnetic fields, but also is an efficient engine that directly converts magnetic energy into particle energies in plasmas \cite{zweibel09a}. 
For comparison, collisionless shocks, regarded to be efficient for particle acceleration in weakly magnetized plasmas, are inefficient in dissipating energy and accelerating non-thermal particles in magnetically dominated flows \cite{sironi15a}. Hence the study of magnetic reconnection in these exotic systems continues to be an interesting topic in high energy astrophysics.

One of the most important issues in relativistic reconnection studies is how fast magnetic energy can be dissipated in the reconnection layer, which determines the time scale of the explosive energy release events. Another related 
problem is the mechanism of non-thermal particle acceleration  \cite{YYuan16a,werner16a,FGuo16a,FGuo15a,FGuo14a,melzani14b,sironi14a,cerutti12a,bessho12a,zenitani01a}. Proposed mechanisms include the direct acceleration by the reconnection electric field at the diffusion region \cite{zenitani01a,uzdensky11a}, the Fermi mechanism at the outflow regions that involves particles bouncing back and forth between reconnection outflows emanated from different x-lines \cite{FGuo14a,dahlin14a,drake06a}, and many other ideas (e.g., \cite{zank14a,drury12a,pino05a}).
In collisionless plasmas, the energy gain of a particle must come from the work done by the electric field $\sim q\int{{\bf E}\cdot {\bf v} dt}$. 
%, where $\bf v$ is the particle velocity. 
Thus, determining the reconnection electric field in the relativistic limit is crucial to determine the acceleration rate and efficiency.

In such magnetically-dominated plasmas, the magnetic energy density is much larger than the rest mass energy density and the Alfv\'en speed approaches the speed of light. Early theoretical work suggested that the magnetic reconnection rate in the relativistic limit may increase compared to the non-relativistic case due to the enhanced inflow arising from the Lorentz contraction of plasma passing through the diffusion region \cite{blackman94a,lyutikov03a}. However, it was later pointed out that the thermal pressure within a pressure-balanced current sheet will constrain the outflow to mildly relativistic conditions, where the Lorentz contraction is negligible \cite{lyubarsky05a} and a relativistic inflow is therefore impossible. 

Recently, fully kinetic simulations by Liu et al.~\cite{yhliu15a} showed that the local inflow speed approaches the speed of light, and the reconnection rate normalized to the immediately upstream condition of the diffusion region can be enhanced to $\sim O(1)$ in strongly magnetized plasmas. 
%the large-$\sigma$ limit. Here the magnetization parameter $\sigma=B^2/(4\pi w)$ is the ratio of magnetic energy density to enthalpy ($w$). 
%Other similar reports includes the simulation using a two-fluid model \cite{zenitani}, in low-$\beta$ collisionless plasmas \cite{bessho}. 
However, the global reconnection rate normalized to the far upstream asymptotic value remains $\lesssim 0.3$ \cite{FGuo15a,yhliu15a,melzani14a,sironi14a,bessho12a,sironi16a} and this discrepancy is not understood. While the relativistic resistive-Petschek model \cite{petschek64a} suggests a similar value for the global rate \cite{lyubarsky05a}, to realize a Petschek solution requires an {\it ad hoc} localized resistivity \cite{biskamp86a,sato79a}, otherwise, the current sheet collapses to the long Sweet-Parker layer \cite{sweet58a,parker57a}. A mechanism for the localized diffusion region is therefore essential to model the reconnection rate. In this Letter, we derive the relation between the global rate and the degree of localization through considering the force balance at the upstream and downstream of the diffusion region. We then propose a mechanism that naturally leads to the localization in such collisionless plasmas.

 {\it Simulation setup--}
The kinetic simulation is performed using a Particle-in-Cell code- VPIC \cite{bowers09a}, which solves the fully relativistic dynamics of particles and electromagnetic fields. The relativistic Harris sheet \cite{yhliu15a,kirk03a,zenitani07a,wliu11a,bessho12a,melzani14a} is employed as the initial condition. The initial magnetic field ${\bf B}=B_{x0} \mbox{tanh}(z/\lambda) \hat{\bf x}$ corresponds to a layer of half-thickness $\lambda$.
Each species has a distribution $f_h \propto \mbox{sech}^2(z/\lambda)\mbox{exp}[-\gamma_d(\gamma_Lmc^2+ mV_d u_y)/T']$ in the simulation frame, which is a component with a peak density $n'_0$ and temperature $T'$ boosted by a drift velocity $\pm V_d$ in the y-direction for positrons and electrons, respectively. In addition, a  non-drifting background component $f_b \propto \mbox{exp}(-\gamma_L m c^2/T_b)$ with a uniform density $n_b$ is included. Here ${\bf u}=\gamma_L {\bf v}$ is the the space-like components of 4-velocity, $\gamma_L=1/[1-(v/c)^2]^{1/2}$ is the Lorentz factor of a particle, and $\gamma_d \equiv 1/[1-(V_d/c)^2]^{1/2}$. The drift velocity is determined by Amp\'ere's law $cB_{x0}/(4\pi\lambda)=2 e\gamma_d n'_0 V_d $.  
The temperature is determined by the pressure balance $B_{x0}^2/(8\pi)=2 n'_0 T'$. The resulting density in the simulation frame is $n_0=\gamma_d n'_0$.
In this Letter, the primed quantities are measured in the fluid rest (proper) frame, while the unprimed quantities are measured in the simulation frame unless otherwise specified.   
Densities are normalized by the initial background density $n_b$, time is normalized by the plasma frequency $\omega_{pe}\equiv(4\pi n_b e^2/m_e)^{1/2}$, velocities are normalized by the light speed $c$, and spatial scales are normalized by the inertial length $d_e\equiv c/\omega_{pe}$. 

The domain size is $L_x\times L_z=384d_e \times 384d_e$ and is resolved by $3072\times6144$ cells. We load 100 macro-particles per cell for each species. The boundary conditions are periodic in the x-direction, while in the z-direction the field boundary condition is conducting and the particles are reflected at the boundaries. The half-thickness of the initial sheet is $\lambda=d_e$, $n_b=n'_0$, $T_b/m_ec^2=0.5$ and $\omega_{pe}/\Omega_{ce}=0.05$ where $\Omega_{ce}\equiv eB_{x0}/(m_e c)$ is a cyclotron frequency. The upstream magnetization parameter is $\sigma_{x0}=B_{x0}^2/(4\pi w)$ with enthalpy $w=2n'_b m_ec^2+[\Gamma/(\Gamma-1)]P'$. Here $\Gamma$ is the ratio of specific heats and $P'\equiv 2n'_b T'_b$ the total thermal pressure. 
For $\Gamma=5/3$ \cite{weinberg72a,synge57a}, $\sigma_{x0}=89$ in this run. A localized perturbation with amplitude $B_z=0.03B_{x0}$ is used to induce a dominant x-line at the center of simulation domain. 

\begin{figure}
\includegraphics[width=8.5cm]{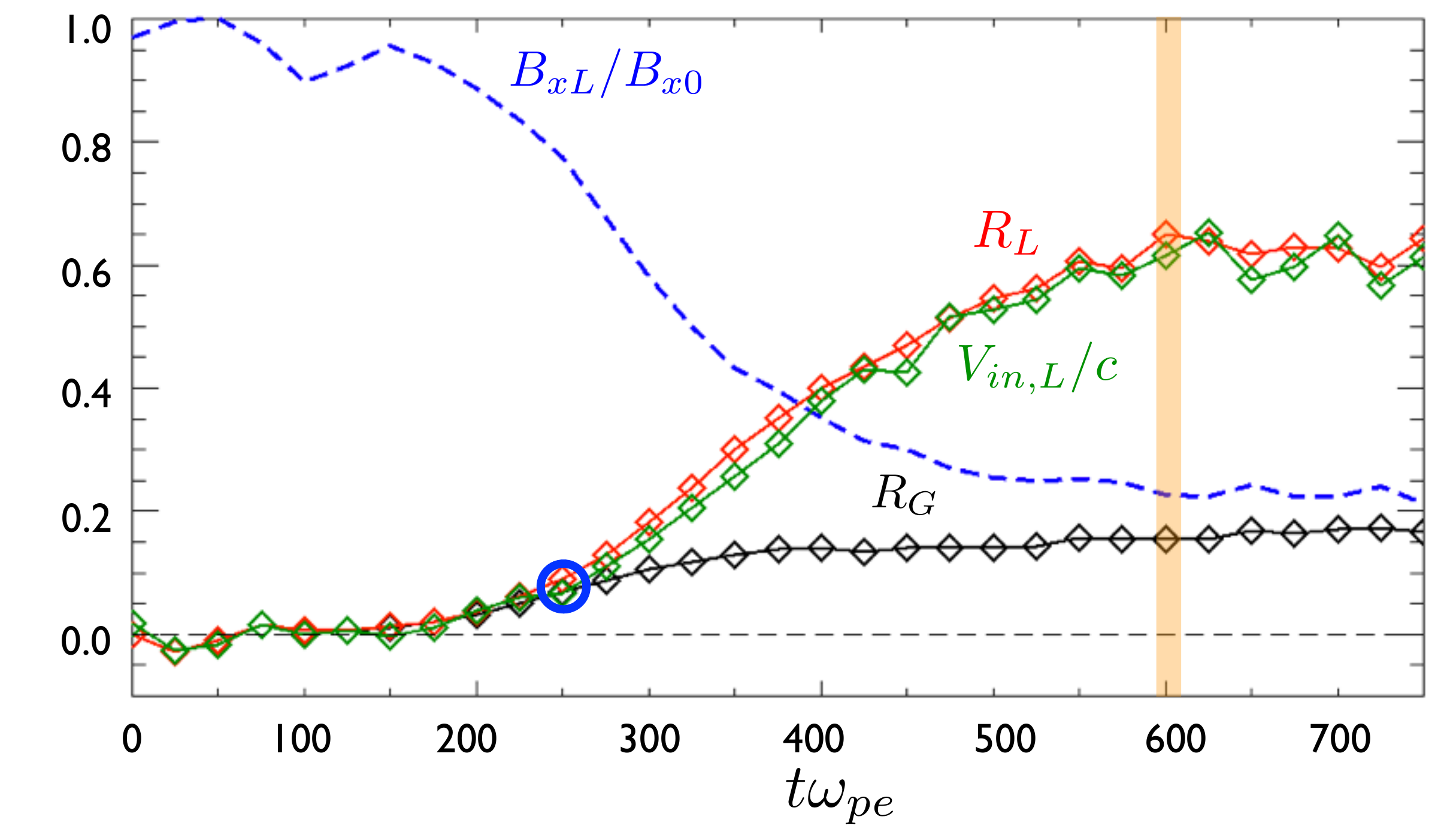} 
\caption {The evolution of measured global reconnection rate $R_G$, local rate $R_L$, local inflow speed $V_{in,L}/c$ and $B_{xL}/B_{x0}$ in a plasma of $\sigma_{x0}=89$. The blue circle marks the deviation of $R_L$ from $R_G$. The grey dashed line at value $0.0$ is for reference. The orange vertical line marks the time for the analyses in Fig.~\ref{feature} and \ref{force_balance}. } 
\label{rate}
\end{figure}

{ \it Simulation results--}
In this Letter, we define the global reconnection rate as $R_G \equiv cE_y/(B_{x0}V_{A0})$
%\begin{equation}
%R_G\equiv \frac{cE_y}{B_{x0}V_{A0}}
%\label{Alfven}
%\end{equation}
and the local reconnection rate as $R_L\equiv cE_y/(B_{xL}V_{AL})$.
%\begin{equation}
%R_L\equiv \frac{cE_y}{B_{xL}V_{AL}}
%\label{Alfven}
%\end{equation}
Subscripts ``0'' and ``L'' indicate quantities far from, and immediately upstream of, the diffusion region where the frozen-in condition ${\bf E}+{\bf V_e}\times {\bf B}=0$ breaks ($|z|\lesssim3.5d_e$ \cite{yhliu15a}). $E_y$ is the reconnection electric field at the x-line and
the Alfv\'en speed in the relativistic limit \cite{sakai80a,anile89a,lichnerowicz67a,yhliu14a} is $V_{A0}=c[\sigma_{x0}/(1+\sigma_{x0})]^{0.5}$ and $V_{AL}=c[\sigma_{xL}/(1+\sigma_{xL})]^{0.5}$ with $\sigma_{xL}\simeq (B_{xL}/B_{x0})^2\sigma_{x0}$.
%\begin{equation}
%V_A=c\sqrt{\frac{\sigma_x}{1+\sigma}}
%\label{Alfven}
%\end{equation}
The evolution of reconnection rates are plotted in Fig.~\ref{rate}, along with the local electron inflow speed, $V_{in,L}$, and the ratio of magnetic fields $B_{xL}/B_{x0}$. Before a quasi-steady state is reached, both the local and global rates increase as the simulation progresses. The deviation of the local rate from the global rate occurs at time $t\simeq 250/\omega_{pe}$ and $B_{xL}/B_{x0}\simeq 0.8$. $R_G$ reaches a plateau of value $\simeq 0.15$ at $t \gtrsim 300/\omega_{pe}$ while $R_L$ continues to grow and $B_{xL}/B_{x0}$ continues to drop. The local rate $R_L$ eventually reaches a plateau of value $\simeq 0.6$ and $B_{xL}/B_{x0}$ reaches a plateau of value $\simeq 0.22$ at time $t\gtrsim 600/\omega_{pe}$. The local inflow speed basically traces the local rate because of the frozen-in condition $E_y\simeq V_{in,L}B_{xL}/c$ and $V_{AL}\simeq c$ in this case, which leads $R_L=V_{in,L}/V_{AL}\simeq V_{in,L}/c$ . The values of these two quantities can approach $\sim O(1)$ with a larger $\sigma_{x0}$, as reported before \cite{yhliu15a,lyutikov16a,zenitani09a}.

\begin{figure}
\includegraphics[width=9cm]{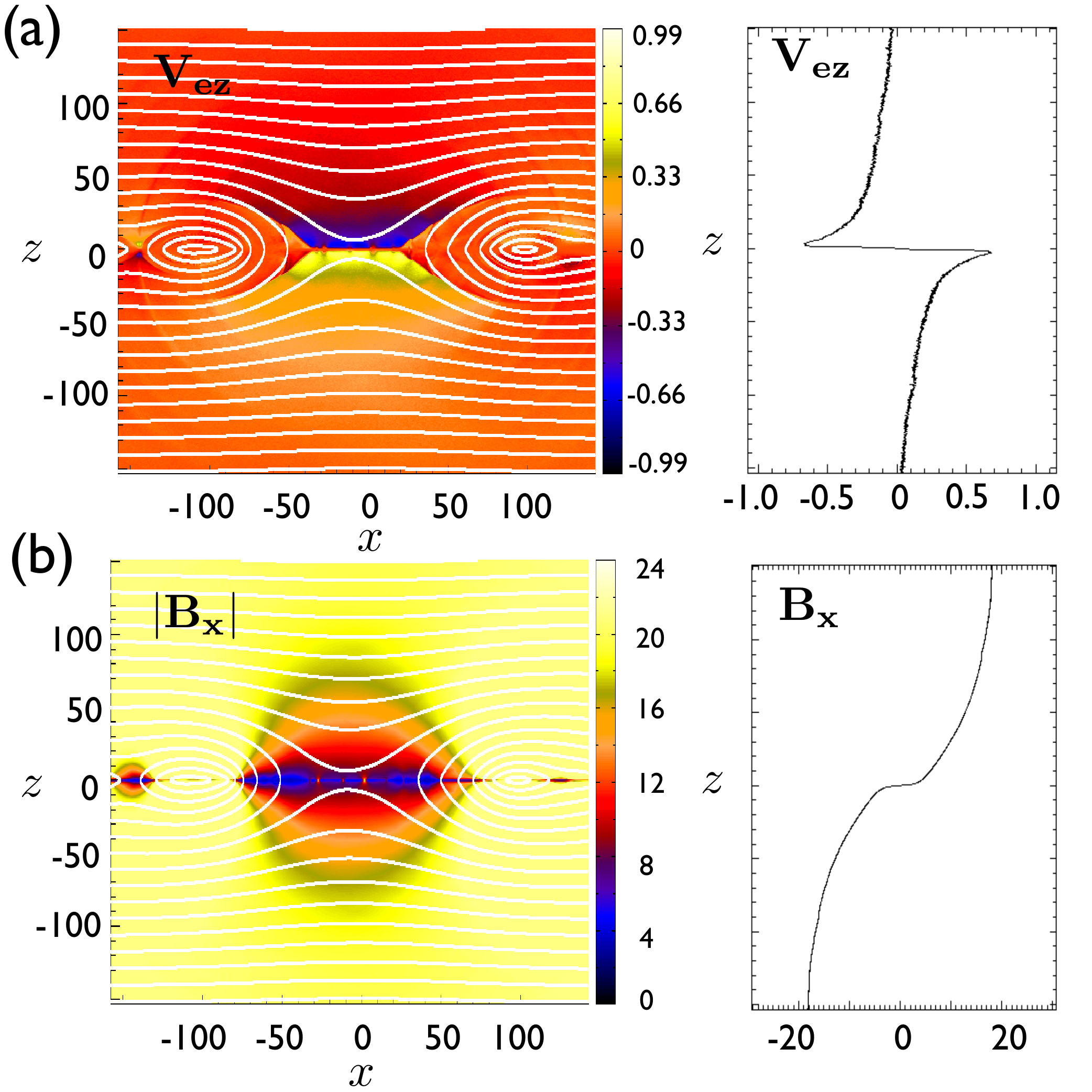} 
\caption {The morphology of relativistic magnetic reconnection at $t=600/\omega_{pe}$. In (a), the $V_{ez}$ and a cut at $x=0$; In (b), the $|B_x|$ and a cut of $B_x$ at $x=0$. The white contour is the in-plane magnetic flux. To better illustrate the variation of the upstream field in (b), we have put an upper limit $B_{x0}$ in the color scale, which artificially reduces the $|B_x|$ around the magnetic islands at outflow exhausts.} 
\label{feature}
\end{figure}

To get a better idea of the spatial variation of the inflow velocity and magnetic fields at the quasi-steady state, the $V_{ez}$ and $B_x$ at time $t=600/\omega_{pe}$ are shown in Fig.~\ref{feature} with the in-plane magnetic flux overlaid. Immediately upstream of the intense thin current sheet, the $|V_{ez}|$ peaks at $|z|\simeq d_e$ with value $\simeq 0.65$, where $B_x$ drops to a value $\simeq 3$. Because of the thin current sheet, $d_e$-scale secondary tearing modes \cite{yhliu15a} are generated repeatedly, which can be seen in Fig.~\ref{feature}. Note that $R_G$ reaches the plateau in Fig.~\ref{rate} long before the generation of secondary tearing modes. 
%Note that in Fig.~\ref{rate}, the ``local'' values are taken at the location where the frozen-in condition ${\bf E}+{\bf V_e}\times {\bf B}=0$ starts to break at the upstream, which is at around $z\simeq 3.5 d_e$ in the nonlinear stage [see the discussion of Fig.~2 in Liu et al. \cite{yhliu15a}]. 
The enhancement of $V_{ez}/c$ closer to the diffusion region is anti-correlated with the reduction of $B_x$ because $E_y\simeq V_{ez} B_x/c$ should be spatially uniform in a quasi-steady state under the 2D constraint, per Faraday's law.

\begin{figure}
\includegraphics[width=8.5cm]{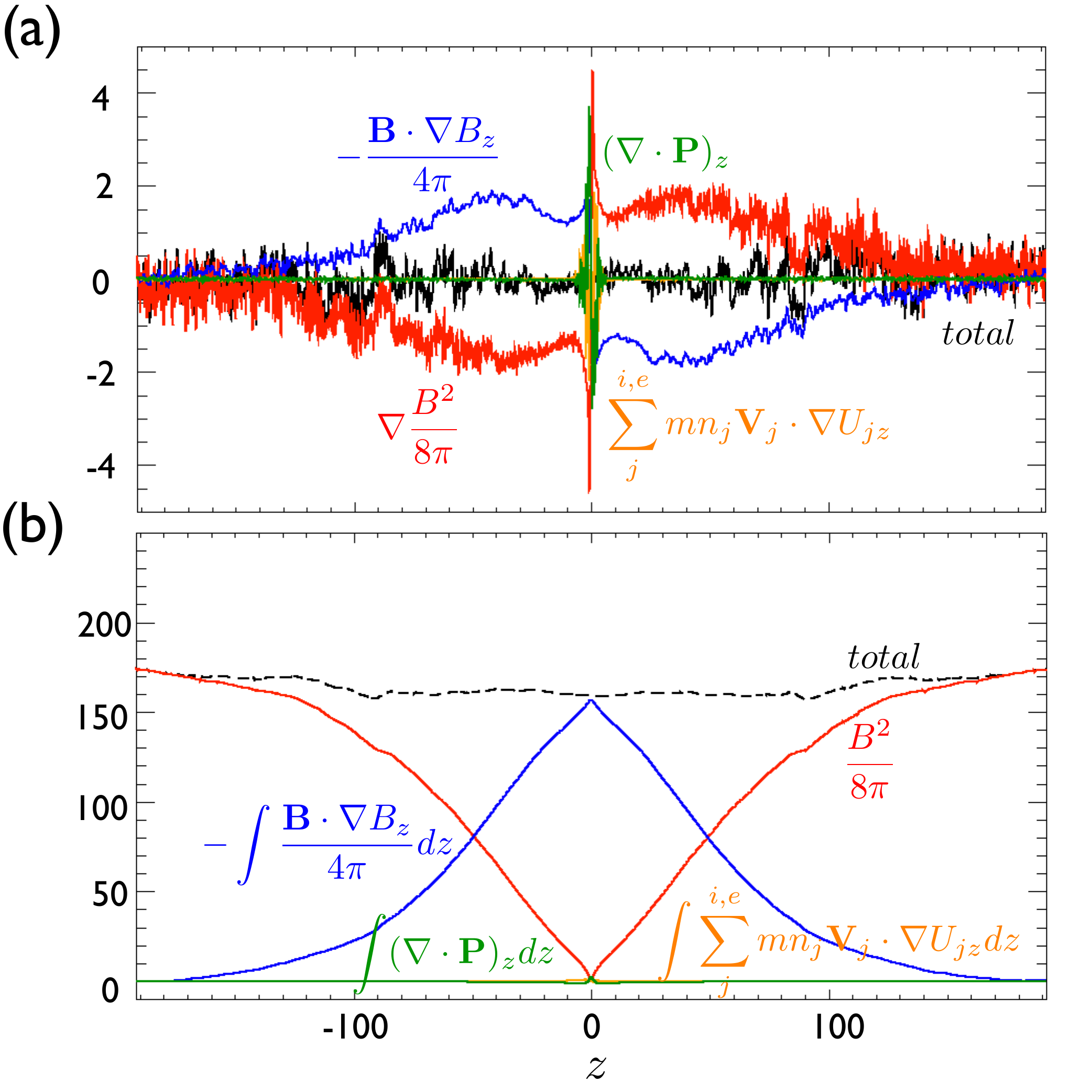} 
\caption {In (a), the force balance in the z-direction along $x=0$ in Fig.~\ref{feature}; In (b), the pressure balance along $x=0$.} 
\label{force_balance}
\end{figure}

To get a clue of how the $B_{xL}$ drops from $B_{x0}$, we examine the force balance across the x-line at $x=0$.
By combining the momentum equations for electrons and positrons \cite{yhliu15a,hesse07a}, the equation of force balance can be derived as %$\sum_j^{e,p}mn_j{\bf V}_j\cdot\nabla{\bf U}_j+\nabla B^2/8\pi+\nabla\cdot {\bf P}-{\bf B}\cdot\nabla {\bf B}/4\pi=-\sum_j^{e,p}mn_j\partial{\bf U}_j/\partial t$.
\begin{equation}
\sum_j^{e,p}mn_j{\bf V}_j\cdot\nabla{\bf U}_j+\nabla\frac{B^2}{8\pi}+\nabla\cdot {\bf P}-\frac{{\bf B}\cdot\nabla {\bf B}}{4\pi}=-\sum_j^{e,p}mn_j\frac{\partial}{\partial t}{\bf U}_j
\label{force}
\end{equation} 
Here the pressure tensor ${\bf P} \equiv \sum_j^{e,p} \int d^3u {\bf v u}f_j-n_j{\bf V}_j{\bf U}_j$, and subscripts ``e'' and ``p'' stand for electrons and positrons respectively. ${\bf U} \equiv (1/n)\int d^3u {\bf u} f$ is the first moment of the space-like components of 4-velocity, and ${\bf V} \equiv (1/n)\int d^3u{\bf v} f$ as usual.
On the left hand side of Eq.~(\ref{force}), the terms represent the inertial force, magnetic pressure gradient force, plasma thermal gradient force and magnetic tension, respectively. In the upstream region the magnetic pressure is balanced by the tension force as shown in Fig.~\ref{force_balance}(a). The thermal pressure is negligible because of the small plasma $\beta\equiv P/(B^2/8\pi)\simeq 0.005$. The time-derivative of inertia is negligible in the quasi-steady state. Therefore, the force balance results in a significant reduction of $B_x$ from the value far upstream at $|z| \gtrsim 150d_e$ to the value immediately upstream of the diffusion region at $|z|\simeq 3.5 d_e$ as shown in the profile of $B^2/8\pi$ in Fig.~\ref{force_balance}(b).

\begin{figure}
\includegraphics[width=8cm]{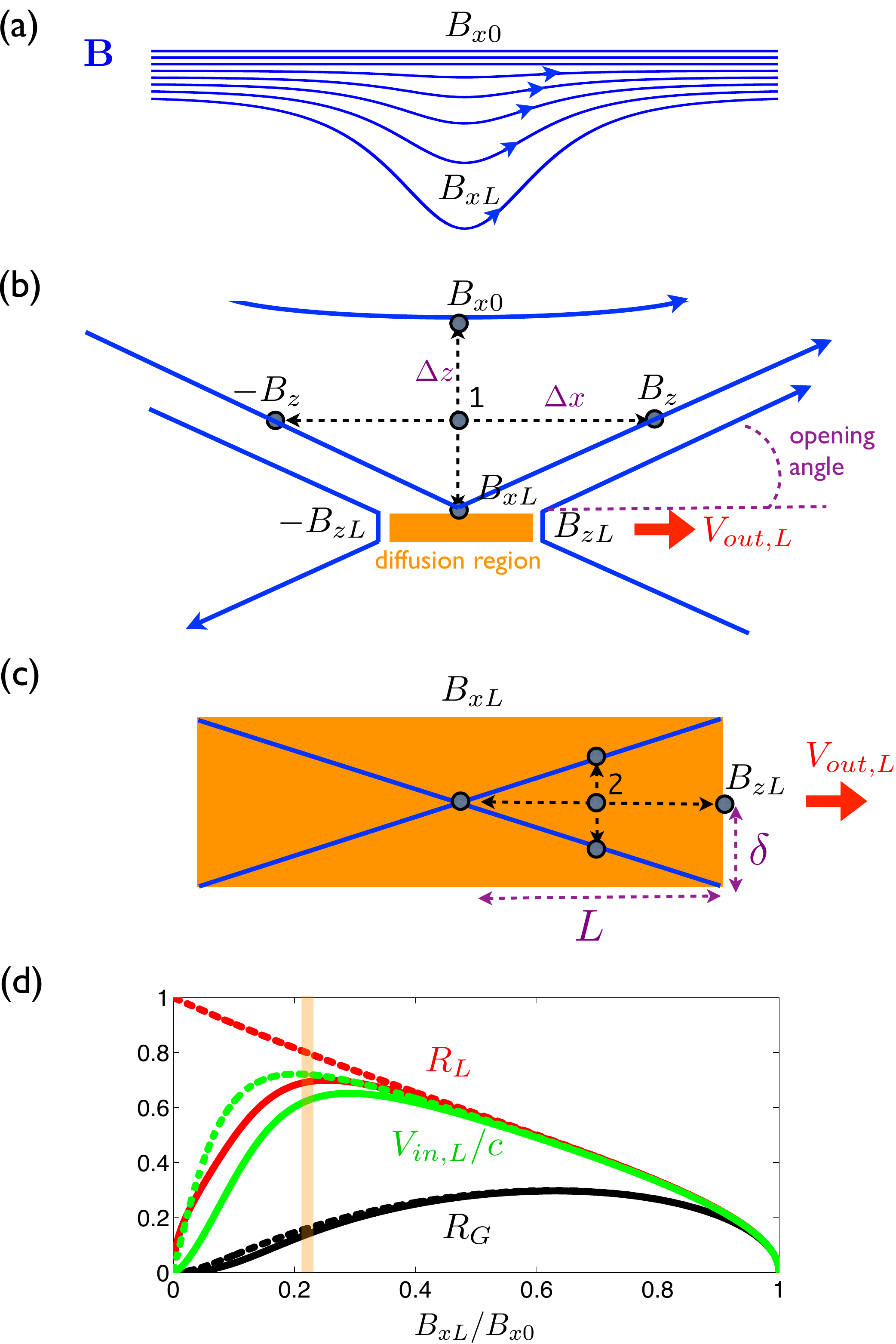} 
\caption {The cartoons of magnetic field lines upstream of the diffusion region ($z >0$) in (a), and the geometry of reconnection in (b). The dimension of the diffusion region in (c). The predictions with $\sigma_{x0}=89$ in (d), the dashed lines use $V_{out,L}=V_{AL}$. The orange vertical line corresponds to that in Fig.~\ref{rate}.} 
\label{model}
\end{figure}

{ \it Simple model--}
When the current sheet pinches locally, it implies a curved upstream magnetic field as illustrated in Fig.~\ref{model}(a). The local magnetic field immediately upstream of the diffusion region, $B_{xL}$, becomes smaller than $B_{x0}$, so that the magnetic pressure gradient force balances the magnetic tension. A larger degree of localization implies a larger curvature, and a smaller $B_{xL}$, as indicated by the ``line-density'' of the in-plane flux in both Fig.~\ref{model}(a), and the upstream region in Fig.~\ref{feature}. Hence, even though the local reconnection rate can be enhanced significantly due to the normalization, the global reconnection rate may not increase much.

To estimate this effect in the $\beta \ll 1$ limit, we analyze the force balance, $\nabla B^2/8\pi \simeq {\bf B}\cdot\nabla {\bf B}/4\pi$, at point 1 marked in Fig.~\ref{model}(b): 
%$nm(V_{z0}^2-V_{zL}^2)/\Delta z+(B_{x0}^2-B_{xL}^2)/8\pi\Delta z\simeq [(B_{x0}+B_{xL})/2](2B_z)/4\pi\Delta x$.
\begin{equation}
\frac{B_{x0}^2-B_{xL}^2}{8\pi\Delta z}\simeq \left(\frac{B_{x0}+B_{xL}}{2}\right)\frac{2B_z}{4\pi\Delta x}.
\label{up_force}
\end{equation}
Note that $\nabla\cdot {\bf B}=0$ is also satisfied at point 1. The $B_x$ at point 1 is linearly interpolated from $B_{x0}$ and $B_{xL}$. %With $E_y\sim V_zB_x/c$, this leads to $(B_{x0}-B_{xL})[1-R_G^2(V_{A0}/V_{A0N})^2(B_{x0}/B_{xL})^2]\simeq 2(\Delta z/\Delta x) B_z$.
The upstream inertial force can be formally ordered out, and it is also negligible in Fig.~\ref{force_balance}.

%Here $V_{A0,n}\equiv B_{x0}/(8\pi n m_e)^{0.5}$ is the non-relativistic definition of Alfv\'en speed in pair plasmas. The $R_G$ term from the upstream inertia force is negligible, as also observed in Fig.~\ref{force_balance}.

A curved upstream magnetic field naturally implies an flaring angle, and that is measured by $\Delta z/\Delta x\simeq B_z/[(B_{x0}+B_{xL})/2]$. For the proof of principle, we match it to the opening angle of the reconnection exhaust just outside of the diffusion region: $\Delta z/\Delta x\simeq B_{zL}/B_{xL}$.
%\begin{equation}
%\frac{\Delta z}{\Delta x}\simeq \frac{B_{zL}}{B_{xL}}.
%\label{Alfven}
%\end{equation} 
We obtain the relation,
\begin{equation}
\frac{B_{zL}}{B_{xL}}\simeq \sqrt{\frac{1-B_{xL}/B_{x0}}{1+B_{xL}/B_{x0}}}.
\label{BzL_BxL}
\end{equation} 
This expression suggests that a larger opening angle requires a further reduction of $B_{xL}/B_{x0}$. In this sense, $B_{xL}/B_{x0}$ gauges the localization of sheet pinch. When $B_{xL}/B_{x0} \rightarrow 0$, the opening angle approaches $45^\circ$ in this model. 

Combined with $E_y\simeq B_{zL} V_{out,L}/c$, the reconnection rates are
\begin{equation}
R_G\simeq \left(\frac{B_{xL}}{B_{x0}}\right)\left(\frac{B_{zL}}{B_{xL}}\right)\left(\frac{V_{out,L}}{V_{A0}}\right);\ R_L\simeq \left(\frac{B_{zL}}{B_{xL}}\right)\left(\frac{V_{out,L}}{V_{AL}}\right)
\label{RG}
\end{equation} 
%\begin{equation}
%R_L\simeq \left(\frac{B_{zL}}{B_{xL}}\right)\left(\frac{V_{out,L}}{V_{AL}}\right)
%\label{RL}
%\end{equation} 
and the local inflow speed is
\begin{equation}
V_{in,L}\simeq R_L V_{AL}.
\label{vin}
\end{equation}

Using Eq.~(\ref{BzL_BxL}) and the outflow speed $V_{out,L}\sim V_{AL}$, the predicted $R_G$, $R_L$ and $V_{in,L}/c$ as functions of $B_{xL}/B_{x0}$ are plotted in Fig.~\ref{model}(d) as dashed-lines.  If $B_{xL}/B_{x0}=1$, the opening angle is zero and reconnection is not expected. In the limit of $B_{xL}/B_{x0}\rightarrow 0$, the reconnecting component vanishes and reconnection ceases (i.e., $R_G=0$). 

However, the geometrical constrain can reduce the outflow speed from $V_{AL}$ when the opening angle approaches $45^\circ$. This correction can be modeled through analyzing the force-balance in the x-direction at point 2 of Fig.~\ref{model}(c): $n'mU_{out}^2/L+B_{zL}^2/8\pi L\simeq (B_{zL}/2)2(B_{xL}/2)/4\pi\delta$, where the inertial force becomes important.
The outflow can be relativistic, $U_{out}\sim \gamma_{out}V_{out,L}\sim V_{out,L}^2/(1-V_{out,L}^2/c^2)$. Assuming the incompressibility of plasmas, then the aspect ratio of the diffusion region $\delta/L\sim B_{zL}/B_{xL}$, and the outflow speed becomes
%\begin{equation}
%\frac{V_{out,L}}{c}\simeq \sqrt{\frac{[1-(B_{zL}/B_{xL})^2](B_{xL}/B_{x0})^2\sigma_{x0}}{1+[1-(B_{zL}/B_{xL})^2](B_{xL}/B_{x0})^2\sigma_{x0}}}
%\label{vout}
%\end{equation} 
\begin{equation}
V_{out,L}\simeq c\sqrt{\frac{(1-B_{zL}^2/B_{xL}^2)\sigma_{xL}}{1+(1-B_{zL}^2/B_{xL}^2)\sigma_{xL}}}.
\label{vout}
\end{equation} 
This expression suggests that when $\delta/L \ll 1$ (i.e., $B_{zL}/B_{xL} \ll 1$) then $V_{out,L}\sim V_{AL}$. When $\delta/L\rightarrow 1$ (i.e., $45^\circ$), the outflow tension is balanced by the magnetic pressure and the outflow vanishes. Plugging Eqs.(\ref{BzL_BxL}) and (\ref{vout}) back to Eqs.~(\ref{RG})-(\ref{vin}), we get the solid curves in Fig.~\ref{model}(d). This correction further constrains the reconnection rate when the opening angle is larger and $B_{xL}/B_{x0}$ is smaller.

This model suggests that during the pinching of the current sheet, a weak localization with $B_{xL}/B_{x0} \lesssim 0.9$ is enough to lead $R_G$ to $\sim 0.2$, then it varies slowly over a wide range of $B_{xL}/B_{x0}$. The local rate $R_L$ and local inflow speed $V_{in,L}/c$ can reach $\sim O(1)$ under stronger localization. The evolution of reconnection rates in Fig.~\ref{rate} can be qualitatively described by this model through decreasing $B_{xL}/B_{x0}$. The rates in the quasi-steady state at time $t=600/\omega_{pe}$ of Fig.~\ref{rate} also compares well with the prediction at $B_{xL}/B_{x0}\simeq 0.22$ with the predicted $R_G \simeq 0.14$, $R_L \simeq 0.69$ and $V_{in,L} \simeq 0.62c$. Given the simplicity of this model, this agreement is quite remarkable. 

%The parameter $B_{xL}/B_{x0}$ remains undetermined in this model. 
While the localization mechanism may vary in different systems, we point out a natural tendency that can lead to the $B_{xL}/B_{x0}$ reduction in such plasmas: A diffusion region sandwiched by a large $B_{xL}\simeq B_{x0}$ at $d_e$-scale (i.e., where the frozen-in condition is broken) requires the current sheet plasma to have a huge thermal pressure to balance the magnetic pressure, and a high drift speed to support the current. For instance, the initial $d_e$-scale current sheet has $T'=100m_ec^2$, $n_0\simeq 10$ and $\gamma_dV_d\simeq 10$. However, the maximum possible reconnection electric field may not be efficient enough in heating and accelerating the cold non-drifting inflowing plasma before they exit the diffusion region \cite{hesse11a}, hence the $B_{xL}$ drops significantly until the $d_e$-scale current sheet becomes sustainable in the quasi-steady state.
If this drop continues with a larger $\sigma_{x0}$, reconnection in the more extreme limit is prone to choke itself off in the quasi-steady state. 
 %$R_G$ may become $\ll 0.1$.  
 %This provides the localization necessary for explaining the $R_G$, $R_L$ and $V_{in,L}$ in the simulation.

%\begin{figure}
%\includegraphics[width=8cm]{model_rate} 
%\caption {The predicted global rate $R_G$, local rate $R_L$ and the local inflow speed $V_{in,L}$ as a function of $B_{xL}/B_{x0}$. In (a) shows the relativistic case corresponding to the simulation; In (b) shows a relativistic case in a more extreme limit; In (c) shows a relativistic case with a guide field; In (d) shows a non-relativistic case. Predictions that include the inertial force are depicted as diamonds.} 
%\label{model_rate}
%\end{figure}

%While the inertial force $\rho {\bf V}\cdot \nabla {\bf U}$ is negligible at upstream ($\sim R_G^2$ smaller), it becomes the dominant term to balances the magnetic tension at downstream to determine the outflow speed $\sim V_{A0}$. While less illuminative, a more sophisticated model that includes inertia force and self-consistently treats the force balance at both the inflow and outflow can be derived. The characteristics of the solution is similar [Appendix].\\

{\it Discussion--}
Knowing the magnitude of electric field is essential for estimating the acceleration of super-thermal particles in highly magnetized astrophysical systems.
This study suggests that the magnitude of the reconnection electric field is bounded by $\sim 30\%$ of the reconnecting component of magnetic field, even in the large-$\sigma_{x0}$ limit. 
While a weak localization of the diffusion region is required, the global reconnection rate $R_G \sim 0.1-0.3$ is not sensitive to a further increase of localization over a wide range of $B_{xL}/B_{x0}$, but the local rate and local inflow speed are. This explains the large difference between the local and global reconnection rates observed in the simulation. 
%Without a guide field, the local inflow velocity $V_{in,L}$ can approach the speed of light when the local $B_x$ becomes small. 

In this model, a larger $\sigma_{x0}$ has little effect on the profile of the global rate $R_G$, but it could make the local inflow speed closer to the speed of light \cite{yhliu15a}. In addition, the effect of a guide field can be included by making the relevant Alfv\'en speed $V_A=c[\sigma_x/(1+\sigma_x+\sigma_g)]^{0.5}$ with $\sigma_g\equiv (B_g/B_{x0})^2\sigma_{x0}$ accounting for the effect a guide field $B_g$. This expression is basically the projection of the total Alfv\'en speed in the outflow direction \cite{yhliu14a,melzani14a,yhliu15a}.  
%The upstream guide field is assumed spatially uniform.  
A guide field also has little effect on $R_G$, but it significantly reduces the local inflow speed and the magnitude of the reconnection electric field through reducing the speed of Alfv\'enic outflows, as observed in Liu et al. \cite{yhliu15a}. The prediction in the non-relativistic and low-$\beta$ limit can be obtained by taking $\sigma_x \ll  1$, and the resulting $R_G$ has a slightly smaller amplitude. \\

\acknowledgments
Y.-H. Liu thanks for helpful discussions with J. Dorelli.  This research was supported by an appointment to the NASA Postdoctoral Program at the NASA-GSFC, administered by Universities Space Research Association through a contract with NASA. Simulations were performed with LANL institutional computing, NASA Advanced Supercomputing and NERSC Advanced Supercomputing.  
%\bibliography{paper}

\begin{thebibliography}{56}
\expandafter\ifx\csname natexlab\endcsname\relax\def\natexlab#1{#1}\fi
\expandafter\ifx\csname bibnamefont\endcsname\relax
  \def\bibnamefont#1{#1}\fi
\expandafter\ifx\csname bibfnamefont\endcsname\relax
  \def\bibfnamefont#1{#1}\fi
\expandafter\ifx\csname citenamefont\endcsname\relax
  \def\citenamefont#1{#1}\fi
\expandafter\ifx\csname url\endcsname\relax
  \def\url#1{\texttt{#1}}\fi
\expandafter\ifx\csname urlprefix\endcsname\relax\def\urlprefix{URL }\fi
\providecommand{\bibinfo}[2]{#2}
\providecommand{\eprint}[2][]{\url{#2}}

\bibitem[{\citenamefont{Coroniti}(1990)}]{coroniti90a}
\bibinfo{author}{\bibfnamefont{F.~V.} \bibnamefont{Coroniti}},
  \bibinfo{journal}{Astrophys. J.} \textbf{\bibinfo{volume}{349}},
  \bibinfo{pages}{538} (\bibinfo{year}{1990}).

\bibitem[{\citenamefont{Arons}(2012)}]{arons12a}
\bibinfo{author}{\bibfnamefont{J.}~\bibnamefont{Arons}},
  \bibinfo{journal}{Space Sci. Rev.} \textbf{\bibinfo{volume}{173}},
  \bibinfo{pages}{341} (\bibinfo{year}{2012}).

\bibitem[{\citenamefont{Lyubarsky and Kirk}(2001)}]{lyubarsky01a}
\bibinfo{author}{\bibfnamefont{Y.}~\bibnamefont{Lyubarsky}} \bibnamefont{and}
  \bibinfo{author}{\bibfnamefont{J.~G.} \bibnamefont{Kirk}},
  \bibinfo{journal}{Astrophys. J.} \textbf{\bibinfo{volume}{547}},
  \bibinfo{pages}{437} (\bibinfo{year}{2001}).

\bibitem[{\citenamefont{{DeVore} et~al.}(2015)\citenamefont{{DeVore},
  Antiochos, Black, Harding, Kalapotharakos, Kazanas, and
  Timokhin}}]{devore15a}
\bibinfo{author}{\bibfnamefont{C.~R.} \bibnamefont{{DeVore}}},
  \bibinfo{author}{\bibfnamefont{S.~K.} \bibnamefont{Antiochos}},
  \bibinfo{author}{\bibfnamefont{C.~E.} \bibnamefont{Black}},
  \bibinfo{author}{\bibfnamefont{A.~K.} \bibnamefont{Harding}},
  \bibinfo{author}{\bibfnamefont{C.}~\bibnamefont{Kalapotharakos}},
  \bibinfo{author}{\bibfnamefont{D.}~\bibnamefont{Kazanas}}, \bibnamefont{and}
  \bibinfo{author}{\bibfnamefont{A.~N.} \bibnamefont{Timokhin}},
  \bibinfo{journal}{Astrophys. J.} \textbf{\bibinfo{volume}{801}},
  \bibinfo{pages}{109} (\bibinfo{year}{2015}).

\bibitem[{\citenamefont{Thompson}(1994)}]{thompson94a}
\bibinfo{author}{\bibfnamefont{C.}~\bibnamefont{Thompson}},
  \bibinfo{journal}{Mon. Not. R. Astron. Soc.} \textbf{\bibinfo{volume}{270}},
  \bibinfo{pages}{480} (\bibinfo{year}{1994}).

\bibitem[{\citenamefont{Zhang and Yan}(2011)}]{zhangB11a}
\bibinfo{author}{\bibfnamefont{B.}~\bibnamefont{Zhang}} \bibnamefont{and}
  \bibinfo{author}{\bibfnamefont{H.}~\bibnamefont{Yan}},
  \bibinfo{journal}{Astrophys. J.} \textbf{\bibinfo{volume}{726}},
  \bibinfo{pages}{90} (\bibinfo{year}{2011}).

\bibitem[{\citenamefont{McKinney and Uzdensky}(2012)}]{mckinney12a}
\bibinfo{author}{\bibfnamefont{J.~C.} \bibnamefont{McKinney}} \bibnamefont{and}
  \bibinfo{author}{\bibfnamefont{D.~A.} \bibnamefont{Uzdensky}},
  \bibinfo{journal}{Mon. Not. R. Astron. Soc.} \textbf{\bibinfo{volume}{419}},
  \bibinfo{pages}{573} (\bibinfo{year}{2012}).

\bibitem[{\citenamefont{Beckwith et~al.}(2008)\citenamefont{Beckwith, Hawley,
  and Krolik}}]{beckwith08a}
\bibinfo{author}{\bibfnamefont{K.}~\bibnamefont{Beckwith}},
  \bibinfo{author}{\bibfnamefont{J.~F.} \bibnamefont{Hawley}},
  \bibnamefont{and} \bibinfo{author}{\bibfnamefont{J.~H.}
  \bibnamefont{Krolik}}, \bibinfo{journal}{Astrophys. J.}
  \textbf{\bibinfo{volume}{678}}, \bibinfo{pages}{1180} (\bibinfo{year}{2008}).

\bibitem[{\citenamefont{Giannios}(2010)}]{giannios10a}
\bibinfo{author}{\bibfnamefont{D.}~\bibnamefont{Giannios}},
  \bibinfo{journal}{Mon. Not. R. Astron. Soc.} \textbf{\bibinfo{volume}{408}},
  \bibinfo{pages}{L46} (\bibinfo{year}{2010}).

\bibitem[{\citenamefont{Jaroschek et~al.}(2004)\citenamefont{Jaroschek, Lesch,
  and Treumann}}]{jaroschek04a}
\bibinfo{author}{\bibfnamefont{C.~H.} \bibnamefont{Jaroschek}},
  \bibinfo{author}{\bibfnamefont{H.}~\bibnamefont{Lesch}}, \bibnamefont{and}
  \bibinfo{author}{\bibfnamefont{R.~A.} \bibnamefont{Treumann}},
  \bibinfo{journal}{Astrophys. J.} \textbf{\bibinfo{volume}{605}},
  \bibinfo{pages}{L9} (\bibinfo{year}{2004}).

\bibitem[{\citenamefont{Abdo et~al.}(2011)\citenamefont{Abdo, Ackermann,
  Ajello, and et~al.}}]{abdo11a}
\bibinfo{author}{\bibfnamefont{A.~A.} \bibnamefont{Abdo}},
  \bibinfo{author}{\bibfnamefont{M.}~\bibnamefont{Ackermann}},
  \bibinfo{author}{\bibfnamefont{M.}~\bibnamefont{Ajello}}, \bibnamefont{and}
  \bibinfo{author}{\bibnamefont{et~al.}}, \bibinfo{journal}{Science}
  \textbf{\bibinfo{volume}{331}}, \bibinfo{pages}{739} (\bibinfo{year}{2011}).

\bibitem[{\citenamefont{Bottcher et~al.}(2013)\citenamefont{Bottcher, Reimer,
  Sweeney, and Parkash}}]{bottcher13a}
\bibinfo{author}{\bibfnamefont{M.}~\bibnamefont{Bottcher}},
  \bibinfo{author}{\bibfnamefont{A.}~\bibnamefont{Reimer}},
  \bibinfo{author}{\bibfnamefont{K.}~\bibnamefont{Sweeney}}, \bibnamefont{and}
  \bibinfo{author}{\bibfnamefont{A.}~\bibnamefont{Parkash}},
  \bibinfo{journal}{Astrophys. J.} \textbf{\bibinfo{volume}{768}},
  \bibinfo{pages}{54} (\bibinfo{year}{2013}).

\bibitem[{\citenamefont{Sironi et~al.}(2015)\citenamefont{Sironi, Petropoulou,
  and Giannios}}]{sironi15a}
\bibinfo{author}{\bibfnamefont{L.}~\bibnamefont{Sironi}},
  \bibinfo{author}{\bibfnamefont{M.}~\bibnamefont{Petropoulou}},
  \bibnamefont{and} \bibinfo{author}{\bibfnamefont{D.}~\bibnamefont{Giannios}},
  \bibinfo{journal}{MNRAS} \textbf{\bibinfo{volume}{450}}, \bibinfo{pages}{183}
  (\bibinfo{year}{2015}).

\bibitem[{\citenamefont{Uhm and Zhang}(2014)}]{uhm14a}
\bibinfo{author}{\bibfnamefont{Z.~L.} \bibnamefont{Uhm}} \bibnamefont{and}
  \bibinfo{author}{\bibfnamefont{B.}~\bibnamefont{Zhang}},
  \bibinfo{journal}{Nature Phys.} \textbf{\bibinfo{volume}{10}},
  \bibinfo{pages}{351} (\bibinfo{year}{2014}).

\bibitem[{\citenamefont{Zweibel and Yamada}(2009)}]{zweibel09a}
\bibinfo{author}{\bibfnamefont{E.~G.} \bibnamefont{Zweibel}} \bibnamefont{and}
  \bibinfo{author}{\bibfnamefont{M.}~\bibnamefont{Yamada}},
  \bibinfo{journal}{ARA\&A} \textbf{\bibinfo{volume}{47}}, \bibinfo{pages}{291}
  (\bibinfo{year}{2009}).

\bibitem[{\citenamefont{Yuan et~al.}(2016)\citenamefont{Yuan, Nalewajko, Zrake,
  East, and Blandford}}]{YYuan16a}
\bibinfo{author}{\bibfnamefont{Y.}~\bibnamefont{Yuan}},
  \bibinfo{author}{\bibfnamefont{K.}~\bibnamefont{Nalewajko}},
  \bibinfo{author}{\bibfnamefont{J.}~\bibnamefont{Zrake}},
  \bibinfo{author}{\bibfnamefont{W.~E.} \bibnamefont{East}}, \bibnamefont{and}
  \bibinfo{author}{\bibfnamefont{R.~D.} \bibnamefont{Blandford}},
  \bibinfo{journal}{arXiv:1604.03179v1}  (\bibinfo{year}{2016}).

\bibitem[{\citenamefont{Werner et~al.}(2016)\citenamefont{Werner, Uzdensky,
  Cerutti, Nalewajko, and Begelman}}]{werner16a}
\bibinfo{author}{\bibfnamefont{G.~R.} \bibnamefont{Werner}},
  \bibinfo{author}{\bibfnamefont{D.~A.} \bibnamefont{Uzdensky}},
  \bibinfo{author}{\bibfnamefont{B.}~\bibnamefont{Cerutti}},
  \bibinfo{author}{\bibfnamefont{K.}~\bibnamefont{Nalewajko}},
  \bibnamefont{and} \bibinfo{author}{\bibfnamefont{M.~C.}
  \bibnamefont{Begelman}}, \bibinfo{journal}{Astrophys. J. Lett.}
  \textbf{\bibinfo{volume}{816}}, \bibinfo{pages}{L8} (\bibinfo{year}{2016}).

\bibitem[{\citenamefont{Guo et~al.}(2016)\citenamefont{Guo, Li, Li, Daughton,
  Zhang, {Lloyd-Ronning}, {Yi-Hsin Liu}, Zhang, and Deng}}]{FGuo16a}
\bibinfo{author}{\bibfnamefont{F.}~\bibnamefont{Guo}},
  \bibinfo{author}{\bibfnamefont{X.}~\bibnamefont{Li}},
  \bibinfo{author}{\bibfnamefont{H.}~\bibnamefont{Li}},
  \bibinfo{author}{\bibfnamefont{W.}~\bibnamefont{Daughton}},
  \bibinfo{author}{\bibfnamefont{B.}~\bibnamefont{Zhang}},
  \bibinfo{author}{\bibfnamefont{N.}~\bibnamefont{{Lloyd-Ronning}}},
  \bibinfo{author}{\bibnamefont{{Yi-Hsin Liu}}},
  \bibinfo{author}{\bibfnamefont{H.}~\bibnamefont{Zhang}}, \bibnamefont{and}
  \bibinfo{author}{\bibfnamefont{W.}~\bibnamefont{Deng}},
  \bibinfo{journal}{Astrophys. J. Lett.} \textbf{\bibinfo{volume}{818}},
  \bibinfo{pages}{L9} (\bibinfo{year}{2016}).

\bibitem[{\citenamefont{Guo et~al.}(2015)\citenamefont{Guo, {Yi-Hsin Liu},
  Daughton, and Li}}]{FGuo15a}
\bibinfo{author}{\bibfnamefont{F.}~\bibnamefont{Guo}},
  \bibinfo{author}{\bibnamefont{{Yi-Hsin Liu}}},
  \bibinfo{author}{\bibfnamefont{W.}~\bibnamefont{Daughton}}, \bibnamefont{and}
  \bibinfo{author}{\bibfnamefont{H.}~\bibnamefont{Li}},
  \bibinfo{journal}{Astrophys. J.} \textbf{\bibinfo{volume}{806}},
  \bibinfo{pages}{167} (\bibinfo{year}{2015}).

\bibitem[{\citenamefont{Guo et~al.}(2014)\citenamefont{Guo, Li, Daughton, and
  {Yi-Hsin Liu}}}]{FGuo14a}
\bibinfo{author}{\bibfnamefont{F.}~\bibnamefont{Guo}},
  \bibinfo{author}{\bibfnamefont{H.}~\bibnamefont{Li}},
  \bibinfo{author}{\bibfnamefont{W.}~\bibnamefont{Daughton}}, \bibnamefont{and}
  \bibinfo{author}{\bibnamefont{{Yi-Hsin Liu}}}, \bibinfo{journal}{Phys. Rev.
  Lett.} \textbf{\bibinfo{volume}{113}}, \bibinfo{pages}{155005}
  (\bibinfo{year}{2014}).

\bibitem[{\citenamefont{Melzani
  et~al.}(2014{\natexlab{a}})\citenamefont{Melzani, Walder, Folini,
  Winisdoerfer, and Favre}}]{melzani14b}
\bibinfo{author}{\bibfnamefont{M.}~\bibnamefont{Melzani}},
  \bibinfo{author}{\bibfnamefont{R.}~\bibnamefont{Walder}},
  \bibinfo{author}{\bibfnamefont{D.}~\bibnamefont{Folini}},
  \bibinfo{author}{\bibfnamefont{C.}~\bibnamefont{Winisdoerfer}},
  \bibnamefont{and} \bibinfo{author}{\bibfnamefont{J.~M.} \bibnamefont{Favre}},
  \bibinfo{journal}{A \& A} \textbf{\bibinfo{volume}{570}},
  \bibinfo{pages}{A112} (\bibinfo{year}{2014}{\natexlab{a}}).

\bibitem[{\citenamefont{Sironi and Spitkovsky}(2014)}]{sironi14a}
\bibinfo{author}{\bibfnamefont{L.}~\bibnamefont{Sironi}} \bibnamefont{and}
  \bibinfo{author}{\bibfnamefont{A.}~\bibnamefont{Spitkovsky}},
  \bibinfo{journal}{Astrophys. J.} \textbf{\bibinfo{volume}{783}},
  \bibinfo{pages}{L21} (\bibinfo{year}{2014}).

\bibitem[{\citenamefont{Cerutti et~al.}(2012)\citenamefont{Cerutti, Werner,
  Uzdensky, and Begelman}}]{cerutti12a}
\bibinfo{author}{\bibfnamefont{B.}~\bibnamefont{Cerutti}},
  \bibinfo{author}{\bibfnamefont{G.~R.} \bibnamefont{Werner}},
  \bibinfo{author}{\bibfnamefont{D.~A.} \bibnamefont{Uzdensky}},
  \bibnamefont{and} \bibinfo{author}{\bibfnamefont{M.~C.}
  \bibnamefont{Begelman}}, \bibinfo{journal}{Astrophys. J. Lett.}
  \textbf{\bibinfo{volume}{754}}, \bibinfo{pages}{L33} (\bibinfo{year}{2012}).

\bibitem[{\citenamefont{Bessho and Bhattacharjee}(2012)}]{bessho12a}
\bibinfo{author}{\bibfnamefont{N.}~\bibnamefont{Bessho}} \bibnamefont{and}
  \bibinfo{author}{\bibfnamefont{A.}~\bibnamefont{Bhattacharjee}},
  \bibinfo{journal}{Astrophys. J.} \textbf{\bibinfo{volume}{750}},
  \bibinfo{pages}{129} (\bibinfo{year}{2012}).

\bibitem[{\citenamefont{Zenitani and Hoshino}(2001)}]{zenitani01a}
\bibinfo{author}{\bibfnamefont{S.}~\bibnamefont{Zenitani}} \bibnamefont{and}
  \bibinfo{author}{\bibfnamefont{H.}~\bibnamefont{Hoshino}},
  \bibinfo{journal}{Astrophys. J.} \textbf{\bibinfo{volume}{562}},
  \bibinfo{pages}{L63} (\bibinfo{year}{2001}).

\bibitem[{\citenamefont{Uzdensity et~al.}(2011)\citenamefont{Uzdensity,
  Cerutti, and Begelman}}]{uzdensky11a}
\bibinfo{author}{\bibfnamefont{D.~A.} \bibnamefont{Uzdensity}},
  \bibinfo{author}{\bibfnamefont{B.}~\bibnamefont{Cerutti}}, \bibnamefont{and}
  \bibinfo{author}{\bibfnamefont{M.~C.} \bibnamefont{Begelman}},
  \bibinfo{journal}{Astrophys. J. Lett.} \textbf{\bibinfo{volume}{737}},
  \bibinfo{pages}{L40} (\bibinfo{year}{2011}).

\bibitem[{\citenamefont{Dahlin et~al.}(2014)\citenamefont{Dahlin, Drake, and
  Swisdak}}]{dahlin14a}
\bibinfo{author}{\bibfnamefont{J.~T.} \bibnamefont{Dahlin}},
  \bibinfo{author}{\bibfnamefont{J.~F.} \bibnamefont{Drake}}, \bibnamefont{and}
  \bibinfo{author}{\bibfnamefont{M.}~\bibnamefont{Swisdak}},
  \bibinfo{journal}{Phys. Plasmas} \textbf{\bibinfo{volume}{21}},
  \bibinfo{pages}{092304} (\bibinfo{year}{2014}).

\bibitem[{\citenamefont{Drake et~al.}(2006)\citenamefont{Drake, Swisdak, Che,
  and Shay}}]{drake06a}
\bibinfo{author}{\bibfnamefont{J.~F.} \bibnamefont{Drake}},
  \bibinfo{author}{\bibfnamefont{M.}~\bibnamefont{Swisdak}},
  \bibinfo{author}{\bibfnamefont{H.}~\bibnamefont{Che}}, \bibnamefont{and}
  \bibinfo{author}{\bibfnamefont{M.~A.} \bibnamefont{Shay}},
  \bibinfo{journal}{Nature} \textbf{\bibinfo{volume}{442}},
  \bibinfo{pages}{553} (\bibinfo{year}{2006}).

\bibitem[{\citenamefont{Zank et~al.}(2014)\citenamefont{Zank, {Le Roux}, Webb,
  Dosch, and Khabarova}}]{zank14a}
\bibinfo{author}{\bibfnamefont{G.~P.} \bibnamefont{Zank}},
  \bibinfo{author}{\bibfnamefont{J.~A.} \bibnamefont{{Le Roux}}},
  \bibinfo{author}{\bibfnamefont{G.~M.} \bibnamefont{Webb}},
  \bibinfo{author}{\bibfnamefont{A.}~\bibnamefont{Dosch}}, \bibnamefont{and}
  \bibinfo{author}{\bibfnamefont{O.}~\bibnamefont{Khabarova}},
  \bibinfo{journal}{Astrophys. J.} \textbf{\bibinfo{volume}{797}},
  \bibinfo{pages}{28} (\bibinfo{year}{2014}).

\bibitem[{\citenamefont{Drury}(2012)}]{drury12a}
\bibinfo{author}{\bibfnamefont{L.~O.} \bibnamefont{Drury}},
  \bibinfo{journal}{MNRAS} \textbf{\bibinfo{volume}{422}},
  \bibinfo{pages}{2474} (\bibinfo{year}{2012}).

\bibitem[{\citenamefont{{de Gouveia Dal Pino} and Lazarian}(2005)}]{pino05a}
\bibinfo{author}{\bibfnamefont{E.~M.} \bibnamefont{{de Gouveia Dal Pino}}}
  \bibnamefont{and} \bibinfo{author}{\bibfnamefont{A.}~\bibnamefont{Lazarian}},
  \bibinfo{journal}{A\&A} \textbf{\bibinfo{volume}{441}}, \bibinfo{pages}{845}
  (\bibinfo{year}{2005}).

\bibitem[{\citenamefont{Blackman and Field}(1994)}]{blackman94a}
\bibinfo{author}{\bibfnamefont{E.~G.} \bibnamefont{Blackman}} \bibnamefont{and}
  \bibinfo{author}{\bibfnamefont{G.~B.} \bibnamefont{Field}},
  \bibinfo{journal}{Phys. Rev. Lett.} \textbf{\bibinfo{volume}{72}},
  \bibinfo{pages}{494} (\bibinfo{year}{1994}).

\bibitem[{\citenamefont{Lyutikov and Uzdensky}(2003)}]{lyutikov03a}
\bibinfo{author}{\bibfnamefont{M.}~\bibnamefont{Lyutikov}} \bibnamefont{and}
  \bibinfo{author}{\bibfnamefont{D.}~\bibnamefont{Uzdensky}},
  \bibinfo{journal}{Astrophys. J.} \textbf{\bibinfo{volume}{589}},
  \bibinfo{pages}{893} (\bibinfo{year}{2003}).

\bibitem[{\citenamefont{Lyubarsky}(2005)}]{lyubarsky05a}
\bibinfo{author}{\bibfnamefont{Y.~E.} \bibnamefont{Lyubarsky}},
  \bibinfo{journal}{MNRAS} \textbf{\bibinfo{volume}{358}}, \bibinfo{pages}{113}
  (\bibinfo{year}{2005}).

\bibitem[{\citenamefont{{Yi-Hsin Liu} et~al.}(2015)\citenamefont{{Yi-Hsin Liu},
  Guo, Daughton, Li, and Hesse}}]{yhliu15a}
\bibinfo{author}{\bibnamefont{{Yi-Hsin Liu}}},
  \bibinfo{author}{\bibfnamefont{F.}~\bibnamefont{Guo}},
  \bibinfo{author}{\bibfnamefont{W.}~\bibnamefont{Daughton}},
  \bibinfo{author}{\bibfnamefont{H.}~\bibnamefont{Li}}, \bibnamefont{and}
  \bibinfo{author}{\bibfnamefont{M.}~\bibnamefont{Hesse}},
  \bibinfo{journal}{Phys. Rev. Lett.} \textbf{\bibinfo{volume}{114}},
  \bibinfo{pages}{095002} (\bibinfo{year}{2015}).

\bibitem[{\citenamefont{Melzani
  et~al.}(2014{\natexlab{b}})\citenamefont{Melzani, Walder, Folini,
  Winisdoerfer, and Favre}}]{melzani14a}
\bibinfo{author}{\bibfnamefont{M.}~\bibnamefont{Melzani}},
  \bibinfo{author}{\bibfnamefont{R.}~\bibnamefont{Walder}},
  \bibinfo{author}{\bibfnamefont{D.}~\bibnamefont{Folini}},
  \bibinfo{author}{\bibfnamefont{C.}~\bibnamefont{Winisdoerfer}},
  \bibnamefont{and} \bibinfo{author}{\bibfnamefont{J.~M.} \bibnamefont{Favre}},
  \bibinfo{journal}{A \& A} \textbf{\bibinfo{volume}{570}},
  \bibinfo{pages}{A111} (\bibinfo{year}{2014}{\natexlab{b}}).

\bibitem[{\citenamefont{Sironi et~al.}(2016)\citenamefont{Sironi, Giannios, and
  Petropoulou}}]{sironi16a}
\bibinfo{author}{\bibfnamefont{L.}~\bibnamefont{Sironi}},
  \bibinfo{author}{\bibfnamefont{D.}~\bibnamefont{Giannios}}, \bibnamefont{and}
  \bibinfo{author}{\bibfnamefont{M.}~\bibnamefont{Petropoulou}},
  \bibinfo{journal}{arXiv:1605.02071v1}  (\bibinfo{year}{2016}).

\bibitem[{\citenamefont{Petschek}(1964)}]{petschek64a}
\bibinfo{author}{\bibfnamefont{H.~E.} \bibnamefont{Petschek}}, in
  \emph{\bibinfo{booktitle}{Proc. AAS-NASA Symp. Phys. Solar Flares}}
  (\bibinfo{year}{1964}), vol.~\bibinfo{volume}{50} of
  \emph{\bibinfo{series}{NASA-SP}}, pp. \bibinfo{pages}{425--439}.

\bibitem[{\citenamefont{Biskamp}(1986)}]{biskamp86a}
\bibinfo{author}{\bibfnamefont{D.}~\bibnamefont{Biskamp}},
  \bibinfo{journal}{Phys. Fluids} \textbf{\bibinfo{volume}{29}},
  \bibinfo{pages}{1520} (\bibinfo{year}{1986}).

\bibitem[{\citenamefont{Sato and Hayashi}(1979)}]{sato79a}
\bibinfo{author}{\bibfnamefont{T.}~\bibnamefont{Sato}} \bibnamefont{and}
  \bibinfo{author}{\bibfnamefont{T.}~\bibnamefont{Hayashi}},
  \bibinfo{journal}{Phys. Fluids} \textbf{\bibinfo{volume}{22}},
  \bibinfo{pages}{1189} (\bibinfo{year}{1979}).

\bibitem[{\citenamefont{Sweet}(1958)}]{sweet58a}
\bibinfo{author}{\bibfnamefont{P.~A.} \bibnamefont{Sweet}}, in
  \emph{\bibinfo{booktitle}{IAU Symp. in Electromagnetic Phenomena in Cosmical
  Physics, ed. B. Lehnet (New York: Cambridge Univ. Press)}}
  (\bibinfo{year}{1958}), p. \bibinfo{pages}{123}.

\bibitem[{\citenamefont{Parker}(1957)}]{parker57a}
\bibinfo{author}{\bibfnamefont{E.~N.} \bibnamefont{Parker}},
  \bibinfo{journal}{J. Geophys. Res.} \textbf{\bibinfo{volume}{62}},
  \bibinfo{pages}{509} (\bibinfo{year}{1957}).

\bibitem[{\citenamefont{Bowers et~al.}(2009)\citenamefont{Bowers, Albright,
  Yin, Daughton, Roytershteyn, Bergen, and Kwan}}]{bowers09a}
\bibinfo{author}{\bibfnamefont{K.}~\bibnamefont{Bowers}},
  \bibinfo{author}{\bibfnamefont{B.}~\bibnamefont{Albright}},
  \bibinfo{author}{\bibfnamefont{L.}~\bibnamefont{Yin}},
  \bibinfo{author}{\bibfnamefont{W.}~\bibnamefont{Daughton}},
  \bibinfo{author}{\bibfnamefont{V.}~\bibnamefont{Roytershteyn}},
  \bibinfo{author}{\bibfnamefont{B.}~\bibnamefont{Bergen}}, \bibnamefont{and}
  \bibinfo{author}{\bibfnamefont{T.}~\bibnamefont{Kwan}},
  \bibinfo{journal}{Journal of Physics: Conference Series}
  \textbf{\bibinfo{volume}{180}}, \bibinfo{pages}{012055}
  (\bibinfo{year}{2009}).

\bibitem[{\citenamefont{Kirk and Skjeraasen}(2003)}]{kirk03a}
\bibinfo{author}{\bibfnamefont{J.~G.} \bibnamefont{Kirk}} \bibnamefont{and}
  \bibinfo{author}{\bibfnamefont{O.}~\bibnamefont{Skjeraasen}},
  \bibinfo{journal}{Astrophys. J.} \textbf{\bibinfo{volume}{591}},
  \bibinfo{pages}{366} (\bibinfo{year}{2003}).

\bibitem[{\citenamefont{Zenitani and Hoshino}(2007)}]{zenitani07a}
\bibinfo{author}{\bibfnamefont{S.}~\bibnamefont{Zenitani}} \bibnamefont{and}
  \bibinfo{author}{\bibfnamefont{H.}~\bibnamefont{Hoshino}},
  \bibinfo{journal}{Astrophys. J.} \textbf{\bibinfo{volume}{670}},
  \bibinfo{pages}{702} (\bibinfo{year}{2007}).

\bibitem[{\citenamefont{Liu et~al.}(2011)\citenamefont{Liu, Li, Yin, Albright,
  Bowers, and Liang}}]{wliu11a}
\bibinfo{author}{\bibfnamefont{W.}~\bibnamefont{Liu}},
  \bibinfo{author}{\bibfnamefont{H.}~\bibnamefont{Li}},
  \bibinfo{author}{\bibfnamefont{L.}~\bibnamefont{Yin}},
  \bibinfo{author}{\bibfnamefont{B.~J.} \bibnamefont{Albright}},
  \bibinfo{author}{\bibfnamefont{K.~J.} \bibnamefont{Bowers}},
  \bibnamefont{and} \bibinfo{author}{\bibfnamefont{E.~P.} \bibnamefont{Liang}},
  \bibinfo{journal}{Phys. Plasmas} \textbf{\bibinfo{volume}{18}}
  (\bibinfo{year}{2011}).

\bibitem[{\citenamefont{Weinberg}(1972)}]{weinberg72a}
\bibinfo{author}{\bibfnamefont{S.}~\bibnamefont{Weinberg}},
  \emph{\bibinfo{title}{Gravitation and Cosmology: Principle and Applications
  of the central theory of relativity}} (\bibinfo{publisher}{John Wiley and
  Sons, Inc, New York}, \bibinfo{year}{1972}), chap.~\bibinfo{chapter}{2},
  p.~\bibinfo{pages}{51}.

\bibitem[{\citenamefont{Synge}(1957)}]{synge57a}
\bibinfo{author}{\bibfnamefont{J.~L.} \bibnamefont{Synge}},
  \emph{\bibinfo{title}{The Relativitic Gas}}
  (\bibinfo{publisher}{North-Holland Publication Company, Amsterdam},
  \bibinfo{year}{1957}), chap.~\bibinfo{chapter}{6}, p.~\bibinfo{pages}{60}.

\bibitem[{\citenamefont{Sakai and Kawata}(1980)}]{sakai80a}
\bibinfo{author}{\bibfnamefont{J.}~\bibnamefont{Sakai}} \bibnamefont{and}
  \bibinfo{author}{\bibfnamefont{T.}~\bibnamefont{Kawata}},
  \bibinfo{journal}{J. Phy. Soc. Japan} \textbf{\bibinfo{volume}{49}},
  \bibinfo{pages}{747} (\bibinfo{year}{1980}).

\bibitem[{\citenamefont{Anile}(1989)}]{anile89a}
\bibinfo{author}{\bibfnamefont{A.~M.} \bibnamefont{Anile}},
  \emph{\bibinfo{title}{Relativistic Fluids and Magneto-Fluids}}
  (\bibinfo{publisher}{Cambridge University Press, New York},
  \bibinfo{year}{1989}), chap.~\bibinfo{chapter}{2}, p.~\bibinfo{pages}{34}.

\bibitem[{\citenamefont{Lichnerowicz}(1967)}]{lichnerowicz67a}
\bibinfo{author}{\bibfnamefont{A.}~\bibnamefont{Lichnerowicz}},
  \emph{\bibinfo{title}{Relativistic Hydrodynamics and Magnetohydrodynamics}}
  (\bibinfo{publisher}{W. A. Benjamin Inc., New York}, \bibinfo{year}{1967}),
  chap.~\bibinfo{chapter}{4}, p. \bibinfo{pages}{112}.

\bibitem[{\citenamefont{{Yi-Hsin Liu} et~al.}(2014)\citenamefont{{Yi-Hsin Liu},
  Daughton, Karimabadi, Li, and Gary}}]{yhliu14a}
\bibinfo{author}{\bibnamefont{{Yi-Hsin Liu}}},
  \bibinfo{author}{\bibfnamefont{W.}~\bibnamefont{Daughton}},
  \bibinfo{author}{\bibfnamefont{H.}~\bibnamefont{Karimabadi}},
  \bibinfo{author}{\bibfnamefont{H.}~\bibnamefont{Li}}, \bibnamefont{and}
  \bibinfo{author}{\bibfnamefont{S.~P.} \bibnamefont{Gary}},
  \bibinfo{journal}{Phys. Plasmas} \textbf{\bibinfo{volume}{21}},
  \bibinfo{pages}{022113} (\bibinfo{year}{2014}).

\bibitem[{\citenamefont{Lyutikov et~al.}(2016)\citenamefont{Lyutikov, Sironi,
  Komissarov, and Porth}}]{lyutikov16a}
\bibinfo{author}{\bibfnamefont{M.}~\bibnamefont{Lyutikov}},
  \bibinfo{author}{\bibfnamefont{L.}~\bibnamefont{Sironi}},
  \bibinfo{author}{\bibfnamefont{S.}~\bibnamefont{Komissarov}},
  \bibnamefont{and} \bibinfo{author}{\bibfnamefont{O.}~\bibnamefont{Porth}},
  \bibinfo{journal}{arXiv:1603.05731v1}  (\bibinfo{year}{2016}).

\bibitem[{\citenamefont{Zenitani et~al.}(2009)\citenamefont{Zenitani, Hesse,
  and Kimas}}]{zenitani09a}
\bibinfo{author}{\bibfnamefont{S.}~\bibnamefont{Zenitani}},
  \bibinfo{author}{\bibfnamefont{M.}~\bibnamefont{Hesse}}, \bibnamefont{and}
  \bibinfo{author}{\bibfnamefont{A.}~\bibnamefont{Kimas}},
  \bibinfo{journal}{Astrophys. J.} \textbf{\bibinfo{volume}{696}},
  \bibinfo{pages}{1385} (\bibinfo{year}{2009}).

\bibitem[{\citenamefont{Hesse and Zenitani}(2007)}]{hesse07a}
\bibinfo{author}{\bibfnamefont{M.}~\bibnamefont{Hesse}} \bibnamefont{and}
  \bibinfo{author}{\bibfnamefont{S.}~\bibnamefont{Zenitani}},
  \bibinfo{journal}{Phys. Plasmas} \textbf{\bibinfo{volume}{14}},
  \bibinfo{pages}{112102} (\bibinfo{year}{2007}).

\bibitem[{\citenamefont{Hesse et~al.}(2011)\citenamefont{Hesse, Neukirch,
  Schindler, Kuznetsova, and Zenitani}}]{hesse11a}
\bibinfo{author}{\bibfnamefont{M.}~\bibnamefont{Hesse}},
  \bibinfo{author}{\bibfnamefont{T.}~\bibnamefont{Neukirch}},
  \bibinfo{author}{\bibfnamefont{K.}~\bibnamefont{Schindler}},
  \bibinfo{author}{\bibfnamefont{M.}~\bibnamefont{Kuznetsova}},
  \bibnamefont{and} \bibinfo{author}{\bibfnamefont{S.}~\bibnamefont{Zenitani}},
  \bibinfo{journal}{Space Sci. Rev.} \textbf{\bibinfo{volume}{160}},
  \bibinfo{pages}{3} (\bibinfo{year}{2011}).

\end{thebibliography}

\newpage

\end{document}